\documentclass[twocolumn]{aastex62}
\usepackage{CJK}

\received{March 14, 2019}
\revised{July 16, 2019}
\accepted{July 30, 2019}
\submitjournal{ApJ}

\shorttitle{CROC: Reionization Histories of Present-day Galaxies}
\shortauthors{Zhu, Avestruz, and Gnedin}

\def\lsim{\lesssim}
\def\Msun{{\rm M}_\odot}
\def\xHI{x_{\rm HI}}
\def\xHIth{x_{\rm HI, th}}

\begin{document}
\begin{CJK*}{UTF8}{gkai}
\title{Cosmic Reionization On Computers: Reionization Histories of Present-day Galaxies}

\correspondingauthor{Hanjue Zhu (朱涵珏)}
\email{hanjuezhu@uchicago.edu}
\author[0000-0003-0861-0922]{Hanjue Zhu (朱涵珏)}
\affiliation{The University of Chicago;
Chicago, IL 60637, USA}

\author[0000-0001-8868-0810]{Camille Avestruz}
\affiliation{Kavli Institute for Cosmological Physics;
The University of Chicago;
Chicago, IL 60637, USA}
\affiliation{Enrico Fermi Institute;
The University of Chicago;
Chicago, IL 60637, USA}

\author{Nickolay Y. Gnedin}
\affiliation{Particle Astrophysics Center; 
Fermi National Accelerator Laboratory;
Batavia, IL 60510, USA}
\affiliation{Kavli Institute for Cosmological Physics;
The University of Chicago;
Chicago, IL 60637, USA}
\affiliation{Department of Astronomy \& Astrophysics; 
The University of Chicago; 
Chicago, IL 60637, USA}

\begin{abstract}
We examine the reionization history of present-day galaxies by explicitly tracing the building blocks of halos from the Cosmic Reionization On Computers project.  We track dark matter particles that belong to $z=0$ halos and extract the neutral fractions at corresponding positions during rapid global reionization.  The resulting particle reionization histories allow us to explore different definitions of a halo's reionization redshift and to account for the neutral content of the interstellar medium.  Consistent with previous work, we find a systematic trend of reionization redshift with mass - present day halos with higher masses have earlier reionization times.  Finally, we quantify the spread of reionization times within each halo, which also has a mass dependence.
\end{abstract}


\section{Introduction} \label{sec:intro}

The gas content in the present day intergalactic medium (IGM) between galaxies is highly ionized \citep{gunnandpeterson65}. The history and sources of this ionization is one of the outstanding questions in studies of galaxy formation.

Several current and forthcoming observational probes focus on the origin of IGM ionization, i.e.\ the epoch of \emph{cosmic reionization}.  Currently, the Hubble Space Telescope (HST) detects high-redshift galaxy candidates between $z\sim6$ and $z\sim10$, which are the likely reionization sources \citep{bouwensetal13,mclureetal13,bouwens_etal15,finkelstein_etal15,livermore_etal17,livermore_etal18}.  In the next decade, the James Webb Space Telescope (JWST) will increase the sample of high $z$ galaxies, and future measurements of 21 cm radio emission will probe neutral hydrogen at high redshifts \citep{furlanetto_etal06,moralesandwhyithe_10,robertson_etal15,madau_18}.

One of the interesting topics in reionization studies is the exploration of ionization histories of individual galaxies, including our own galaxy, the Milky Way. For example, cosmic reionization may have left its imprint on the population of dwarf galaxies in the Local Group \citep[for a recent review and earlier references, see][]{bb17,wb17,gbk18,kph18}. In order to investigate the connection between reionization and present-day halos, we need to model halo evolution throughout its entire cosmic history, from the epoch of reionization to the present day. The majority of previous works have examined the reionization history of individual halos using seminumerical methods \citep{alvarez_etal09, busha_etal10,lunnan_etal12,li_etal14}. Most recently, \citet{aubert_etal18} used a fully coupled radiation-hydrodynamics simulation to study the reionization histories of high redshift progenitors of $z=0$ halos from an N-body only simulation, with a focus on halos ranging from $10^8h^{-1}\Msun$ to $10^{13}h^{-1}\Msun$.

In this paper, we use the ``Cosmic Reionization On Computers" (CROC) project to build upon and extend the work of \citet{aubert_etal18}.  CROC produces numerical simulations of reionization that self-consistently model relevant physical processes, providing a physically plausible (though, of course, not perfect) model for cosmic reionization. Our methods and analysis are largely complementary to the work by \citet{aubert_etal18}: we explicitly account for the neutral content of the interstellar medium (ISM) in our definition of reionization redshift, and we look at how reionization varies across the building blocks of present day halos.  We explore in detail the ionization process of a single halo and the very definition of the halo reionization moment.

This paper is organized as follows. Section \ref{sec:methods} briefly describes the CROC simulations and how we calculate the reionization history of halos.  In Section \ref{sec:results}, we first show the variations in reionization history as a function of halo mass at $z=0$ depending on the definition of reionization redshift and quantify the spread in halo reionization times.  Finally, we discuss the overall trend of reionization history with halo mass and future work.

\section{Methodology}\label{sec:methods}
\subsection{CROC Simulations}

\begin{figure*}[t]
    \centering
    \begin{minipage}{0.49\textwidth}
        \centering
        \includegraphics[width=\textwidth]{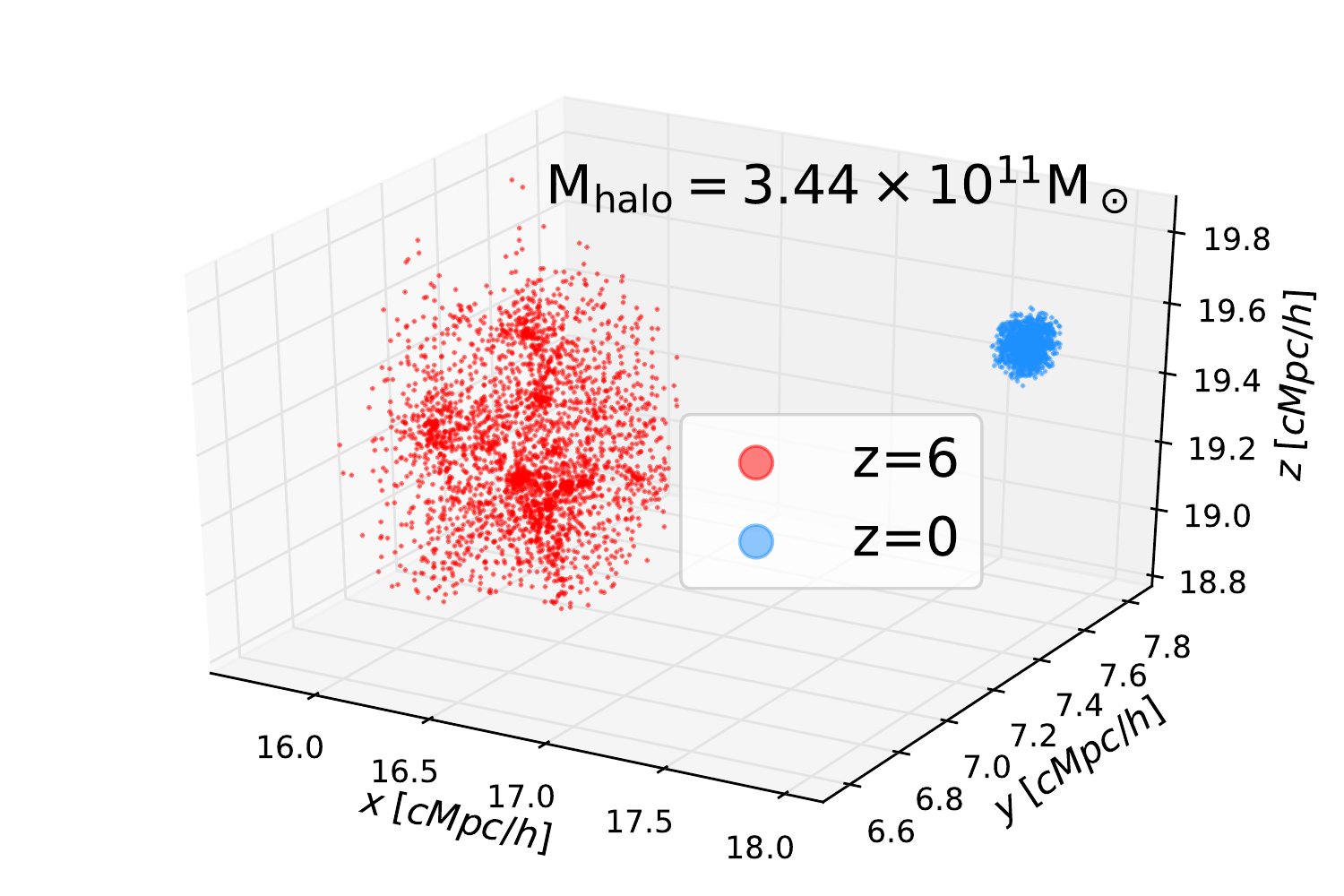} 
    \end{minipage}%
    \begin{minipage}{0.49\textwidth}
        \centering
        \includegraphics[width=\textwidth]{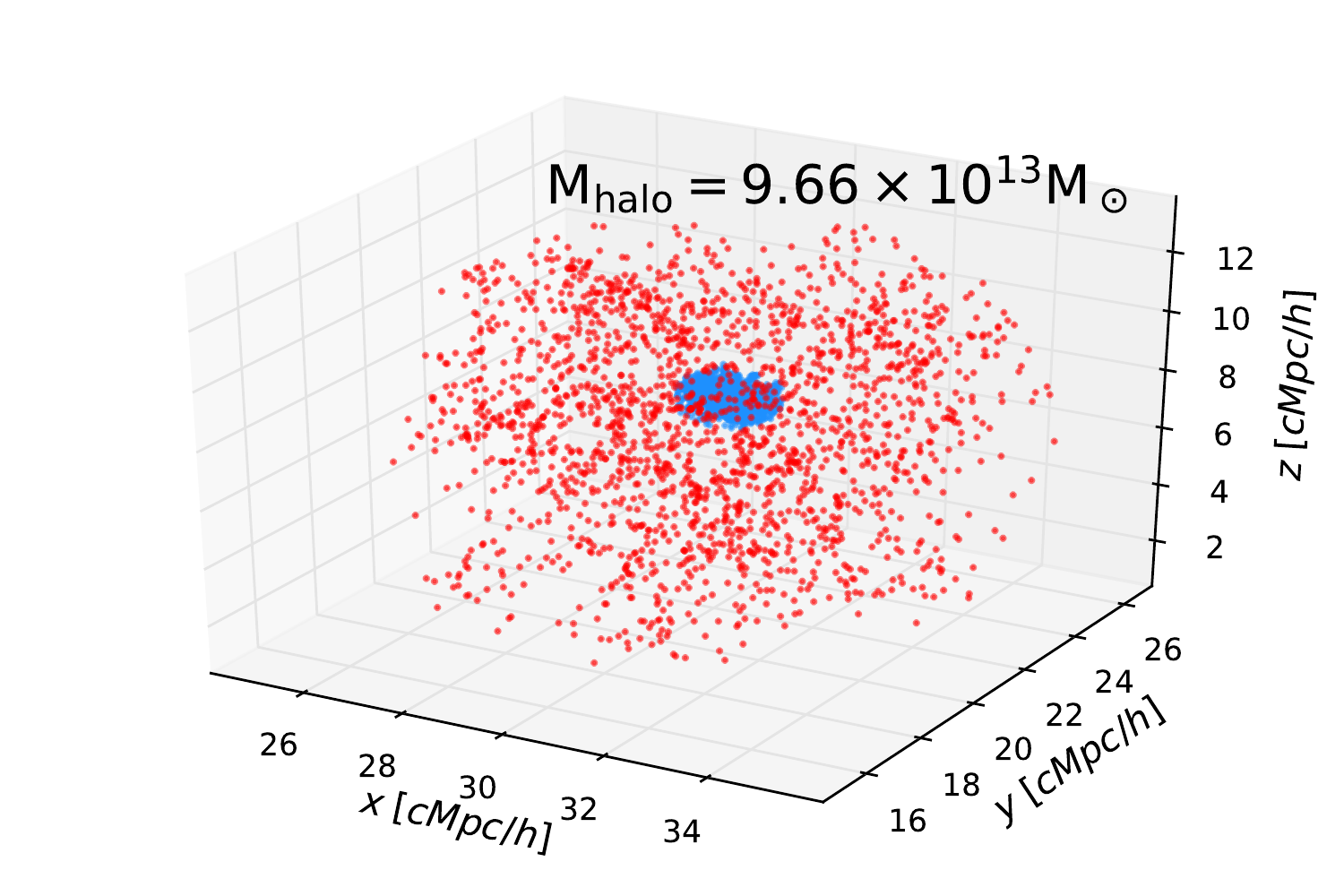} 
    \end{minipage}
\caption{Particle positions of present day halos at $z=0$ and $z=6$ in comoving $h^{-1}$ Mpc. Blue points represent particles at $z=0$; red points represent the same particles at $z=6$.  Two panels from left to right show halos of masses $\approx3\times10^{11}$ and $10^{14}\Msun$ respectively.  The $10^{14}\Msun$ halo is one of the most massive objects in our simulations, and the image shows particles that are downsampled by a factor of 10{,}000.  The $3\times10^{11}\Msun$ halo is downsampled by a factor of 4.\label{fig:positionchange}}
\end{figure*}

All CROC simulations were performed with the Adaptive Refinement Tree (ART) code \citep{kravtsov99,kravtsov_etal02,rudd_etal08}. They include a wide range of physical processes that are thought to be necessary to self-consistently model cosmic reionization, such as gravity and gas dynamics, fully coupled radiative transfer, atomic cooling and heating processes, molecular hydrogen formation, star formation, and stellar feedback. Full details of the simulations are described in the first CROC paper \citep{gnedin14}. 

For our project, we extract data from three independent random realizations of a $40h^{-1}$ comoving Mpc simulation box. Each of the three runs contains $1024^{3}$ dark matter particles and starts with the same number of grid cells. Grid cells are then adaptively refined and unrefined during the simulation, to maintain the approximately Lagrangian behavior (approximately constant gas mass per simulation cell).

These ``full physics" simulations are too expensive to be continued to $z=0$. In order to make a connection to the present day, we completed three dark matter only simulations with the same three sets of initial conditions and the same mass resolution ($1024^3$ particles). We identify halos at $z=0$ with the `yt' implementation of the HOP algorithm \citep{EisensteinandHut_97,yt_11}.  The halo masses extracted by HOP are defined within ${R}_{\rm vir}=R_{160c}$ of the halo center.  This region encompasses an average density that is $\rho_{\rm tot}=160\rho_c$.

\subsection{Defining the Reionization Redshift}\label{sec:methods:reionization_redshift}
The CROC simulations track ionized fractions ($x_{\rm HII}$) and neutral fractions ($\xHI$) in separate fields on the grid.  The global behavior of cosmic reionization in these simulations has already been explored in \citet{gnedinandkaurov_14}.  There, they showed the evolution of mass- and volume-weighted hydrogen fractions with redshift, as well as the distributions of ionized and neutral bubbles at different redshifts. 

To define the reionization redshift of a halo, we use the dark matter particles within $R_{\rm vir}$ of present day halos to trace their progenitors and establish reionization history of the full Lagrangian region of each halo, sampling not only the main progenitor for each halo, but also all resolved satellite halos accreted throughout the whole cosmic history.  

Figure~\ref{fig:positionchange} shows how Lagrangian regions of two different halos change between $z=6$ (in red) and $z=0$ (in blue).  Two panels correspond to halos with $z=0$ masses of $M_{\rm halo}\approx3\times10^{11}\Msun$ and $10^{14}\Msun$ respectively.  Note that the more massive halo collapses from a region of the order of $10h^{-1}$ Mpc in diameter in comoving units.  In defining the reionization history of a halo, we wish to capture the evolution of neutral fraction from this entire Lagrangian region, therefore tracing the full reionization history of each halo.

\begin{figure*} 
\centering
    \centering
    \begin{minipage}{0.49\textwidth}
        \centering
        \includegraphics[width=.95\textwidth]{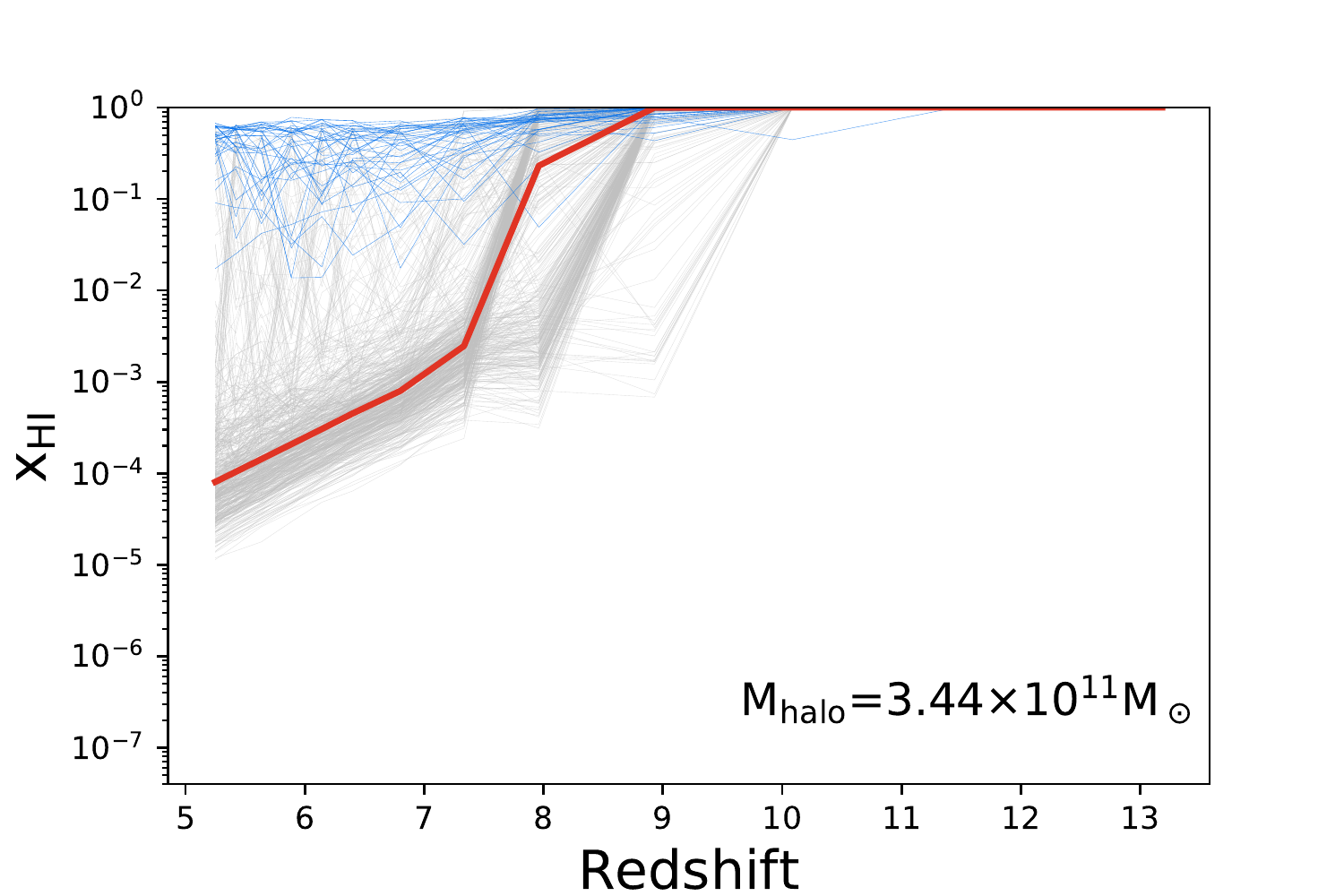}
    \end{minipage}\hfill
    \begin{minipage}{0.49\textwidth}
        \centering
        \includegraphics[width=.95\textwidth]{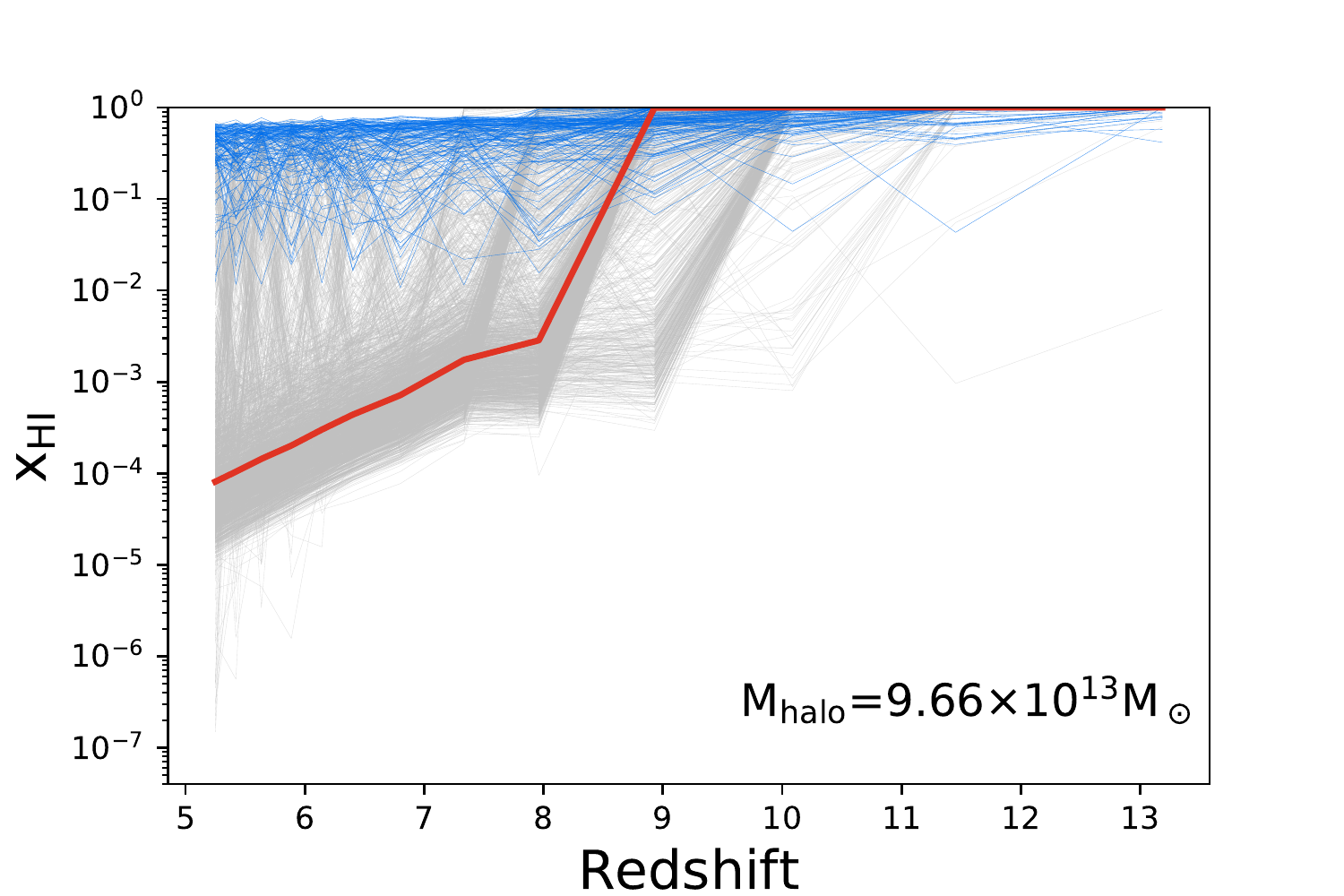}
    \end{minipage}
\caption{Particle reionization histories in two halos from Fig. \ref{fig:positionchange}. Thin grey lines represent individual particles that belong to the $z=0$ halo.  Thin blue lines trace examples of ``ISM" particles for this halo.  The particles from the 3$\times10^{11}\Msun$ halo have been downsampled by a factor of 100 for visualization and the particles from the $10^{14}\Msun$ halo have been downsampled by a factor of 10{,}000.  The median reionization history of all tracer particles of this halo is shown in red.  More massive halos have particles whose reionization histories are more diverse since these halos collapsed from a larger volume.  The median reionization time is also earlier for more massive halos.\label{fig:reionhistory}}
\end{figure*}

We start with the ansatz that the dark matter particles of a halo at $z=0$ trace and sample the gas of halo progenitors. Of course, individual dark matter particles do not trace trajectories of gas elements; only the ensemble of all dark matter particles may provide a sample of the gas properties at any given instant to the extent that these particles sample the density distribution inside the halo. In order to compile the full reionization history of each halo, we track the neutral fraction, $\xHI$, in cells that each dark matter particle passes through, treating the dark matter particles of a $z=0$ halo as tracers.  Each tracer particle therefore has an associated reionization history. 
We show example particle reionization histories in Figure~\ref{fig:reionhistory} for two different halos. Individual particle histories are shown with grey lines, with the red thick line tracing the median particle reionization history within that halo. Note that the more massive halo (right panel) has particles whose reionization histories drop below $10^{-2}$ well before $z=10$, whereas the less massive halo (left panel) does not.  Additionally, the reionization histories of the tracer particles in the more massive halo are more diverse because more massive halos collapse from a larger Lagrangian region.

Operationally, in order to construct reionization history for a given particle, we track its positions in the 15 simulation snapshots we have in the range of $5\leq z\leq 19$\footnote{The reionization history comes from simulation snapshots that are spaced out in time, at $z\approx[5.2,  5.4,  5.7,  5.9,  6.1,
       6.4,  6.8,  7.3,  8.0,  9.0,
      10.0, 11.4, 13.3, 15.6, 18.7]$.  } and find the corresponding neutral fractions of gas in those positions at each snapshot. We then identify the reionization redshift associated with this particle as the moment the particle neutral fraction first falls below a given threshold value, $\xHIth$.

A complicating feature of CROC simulations absent in the previous studies is the existence of a subset of particles that ``never reionize," i.e.\ whose past trajectories are confined to regions that remain significantly neutral (i.e., $\xHI$ always stays above a given threshold) through the entire epoch of reionization ($5\lsim z\lsim 19$). Such particle histories are shown with blue lines in Figure~\ref{fig:reionhistory}. We associate these histories with particles that trace dense gas structures (like the actual galactic disks of progenitor galaxies, which are fully resolved in CROC simulations and which, obviously, always remain neutral) of a $z=0$ halo and label such particles as ``ISM particles." Some of such histories can also be numerical artifacts due to aliasing between particle trajectories and sampling of simulation snapshots. Their reionization histories are not well defined and we exclude such particles from the definition of a ``reionization history of a given halo." 

We define ``ISM" particles as particles whose past trajectories never pass through a spatial region with hydrogen neutral fraction $\xHI$ below $10^{-2}$. We also test other values for the ``ISM threshold" of $\xHI>10^{-1}$ and $\xHI>10^{-3}$, but these do not significantly alter the main results.  We discuss these particles further in Section~\ref{sec:results:neutral_particles}.

We tested two definitions of the reionization redshift of a halo with respect to a chosen neutral fraction threshold, $\xHIth$. The first definition uses the individual particle histories within each halo. The earliest time the neutral fraction along the reionization history of a tracer particle crosses the neutral fraction threshold, $\xHIth$, is our definition of the reionization redshift for that tracer particle.  To calculate this, we smoothly interpolate the particle reionization history, the $\xHI$ values over the redshift snapshots of our simulation. To avoid discretization effects, we further smooth the distribution of reionization redshifts.  We draw a random redshift from the redshift bin whose edges are defined by the simulation snapshots that bracket the earliest neutral fraction crossing time.  This is a reasonable procedure since the true crossing time can be anywhere in that bin.

The halo reionization redshift according to the first definition for a given $\xHIth$ is then simply the median reionization redshift of all of the non-``ISM" tracer particles for that halo.

The second definition uses the median particle history per halo. The thick red line in Figure \ref{fig:reionhistory} illustrates the median history of all particles, excluding the ``ISM" population.  The median of $\xHI$ values at each interpolated redshift defines the median particle reionization history.  A halo reionization redshift for a given $\xHIth$ under this second definition is simply the redshift when the median particle history crosses that threshold.  While the second definition of a halo reionization redshift differs from the first, there is a negligible statistical difference in each mass bin with an rms difference less than 0.5\% in resulting reionization redshift.
However, since we are not considering the reionization history for each individual particle in the second definition, we no longer have the information on the spread of reionization redshifts between particles in the same halo.

We therefore use the first definition to also quantify the spread in particle reionization redshifts and its mass dependence.

\section{Results}\label{sec:results}

\begin{figure*}
    \centering
    \begin{minipage}{0.49\textwidth}
        \centering
        \includegraphics[width=0.99\textwidth]{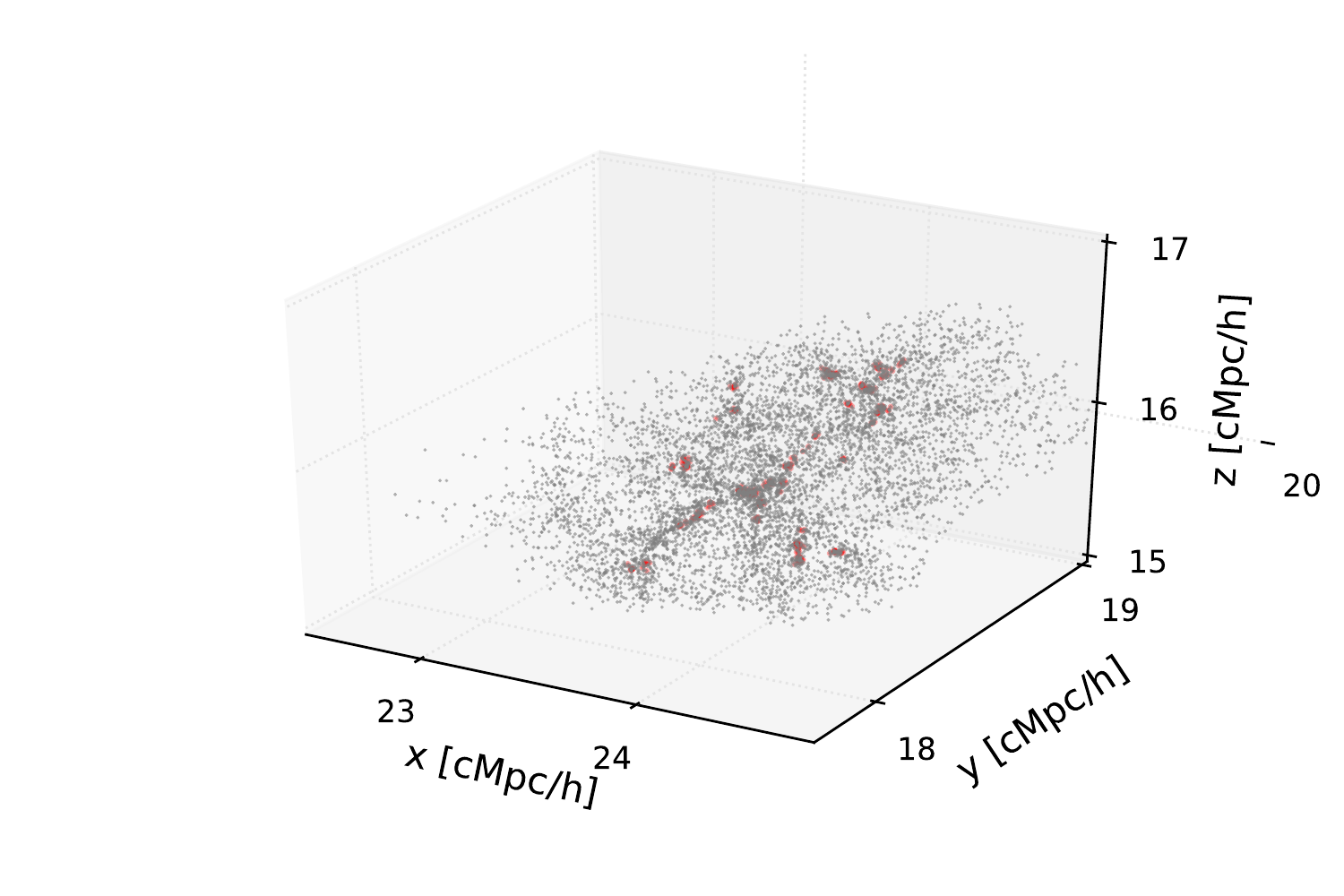} 
    \end{minipage}\hfill
    \begin{minipage}{0.49\textwidth}
        \centering
        \includegraphics[width=0.99\textwidth]{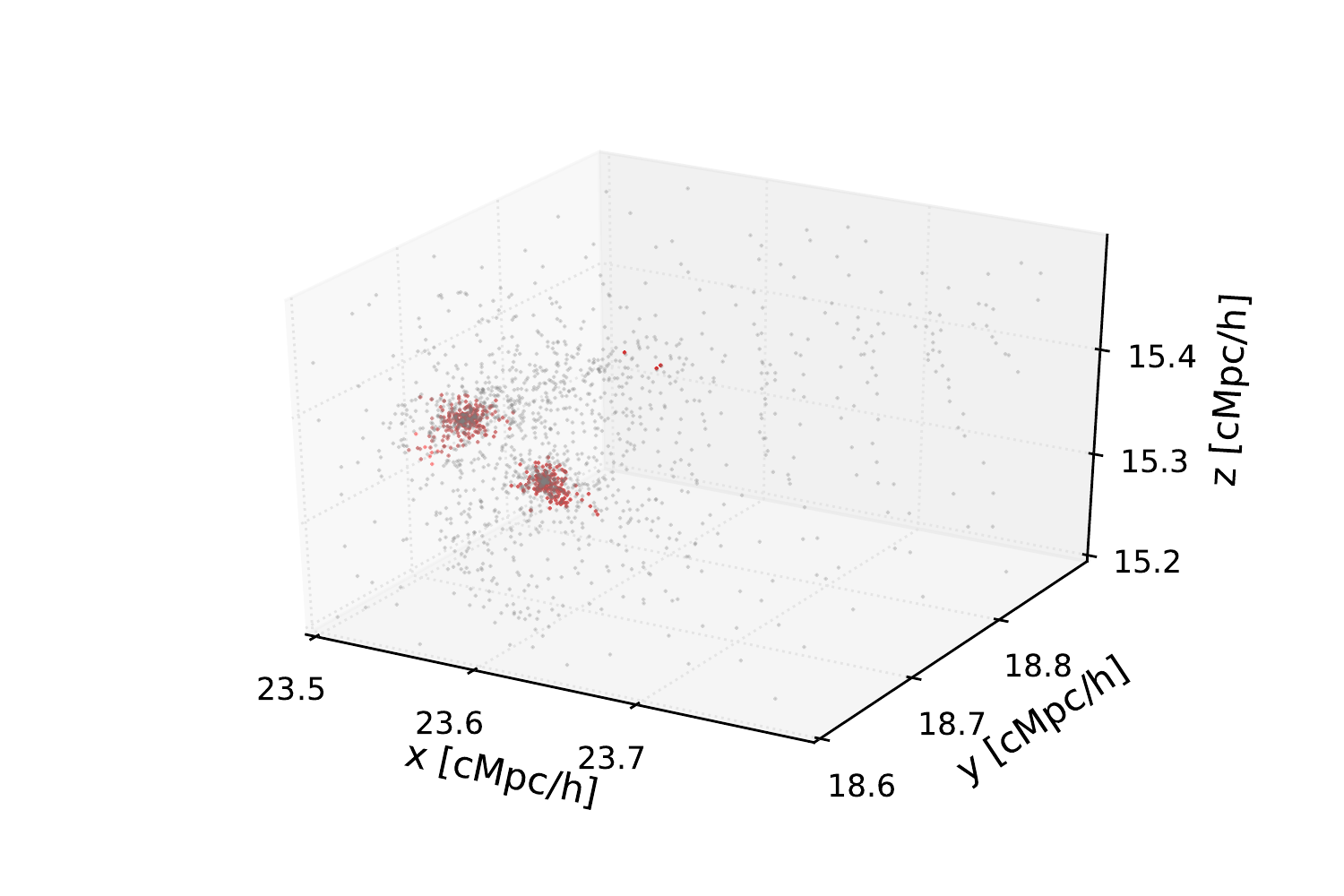} 
\end{minipage}
\caption{Particle positions at $z\approx6$ from a halo of $z=0$ mass $3\times10^{11}\Msun$.  The red points show particles whose reionization history stays relatively neutral through the epoch of reionization (``ISM particles"), staying above a neutral fraction threshold of with $x_{\rm{th,ISM}}>10^{-2}$.  The gray points show all other halo particles. Points have been downsampled by a factor of 8. The right panel shows a zoomed-in view of the particles in two progenitor halos.  Note that decreasing the threshold spreads red points toward more diffuse regions. }\label{fig:neutral_substructure}
\end{figure*}

\subsection{Neutral substructure in halo progenitors}\label{sec:results:neutral_particles}

In computing reionization times for halos, we identify subsets of halo particles whose neutral fraction remains above $x_{\rm{th,ISM}}>10^{-2}$ until $z\approx5$  and refer to these as ``ISM" particles.  We also test the effects of varying the threshold for ``ISM" particles between $x_{\rm{th,ISM}}>10^{-1}$ and $x_{\rm{th,ISM}}>10^{-3}$.  

Figure \ref{fig:neutral_substructure} shows the distribution of progenitor particles at $z=5.89$ for a halo of $z=0$ mass $\approx3\times10^{11}\Msun$. The left panel shows all progenitor particles, and the right panel zooms into a region where we can distinctly see two progenitor halos. In grey, we show the distribution of particles whose associated reionization histories go below the threshold for ``ISM" particles at $z>5$. In red we show the distribution of ``ISM" particles. 

The ``ISM" particle distribution tends to clump in overdense regions at the centers of progenitor halos. These particles trace high local densities that are harder to reionize and reservoirs of molecular hydrogen that continue to form stars. Overdense regions may include the always-neutral galactic disks, damped Lyman-alpha systems, tidal tails from satellites that are analogous to the Magellanic stream in the Milky Way halo, or any other regions that is dense and opaque enough not to be highly ionized by the cosmic background. As noted in Section \ref{sec:methods:reionization_redshift}, we exclude these particles when defining the reionization redshift of a halo and its components.

The choice of $x_{th,ISM}$ makes minimal difference in our main results.  When increasing (decreasing) the threshold defining ``ISM" particles, the ``ISM" particles in Figure~\ref{fig:neutral_substructure} become less (more) concentrated and localized in the halo progenitor centers.  When using a smaller value for the ``ISM" threshold, more dark matter particles are tagged as ``ISM," and extend to more diffuse regions.  However, our main results do not significantly vary across the threshold values we tested, $x_{th,ISM}\in(10^{-1},10^{-2},10^{-3})$. The overall reionization redshift trend with halo mass minimally changes, and the spread of the particle reionization redshifts in each halo systematically decreases.

\subsection{Particle Reionization Redshift Trends with Halo Mass}\label{sec:results:masstrend}

\begin{figure}
    \centering
\begin{minipage}{0.49\textwidth}
        \centering
        \includegraphics[width=0.95\textwidth]{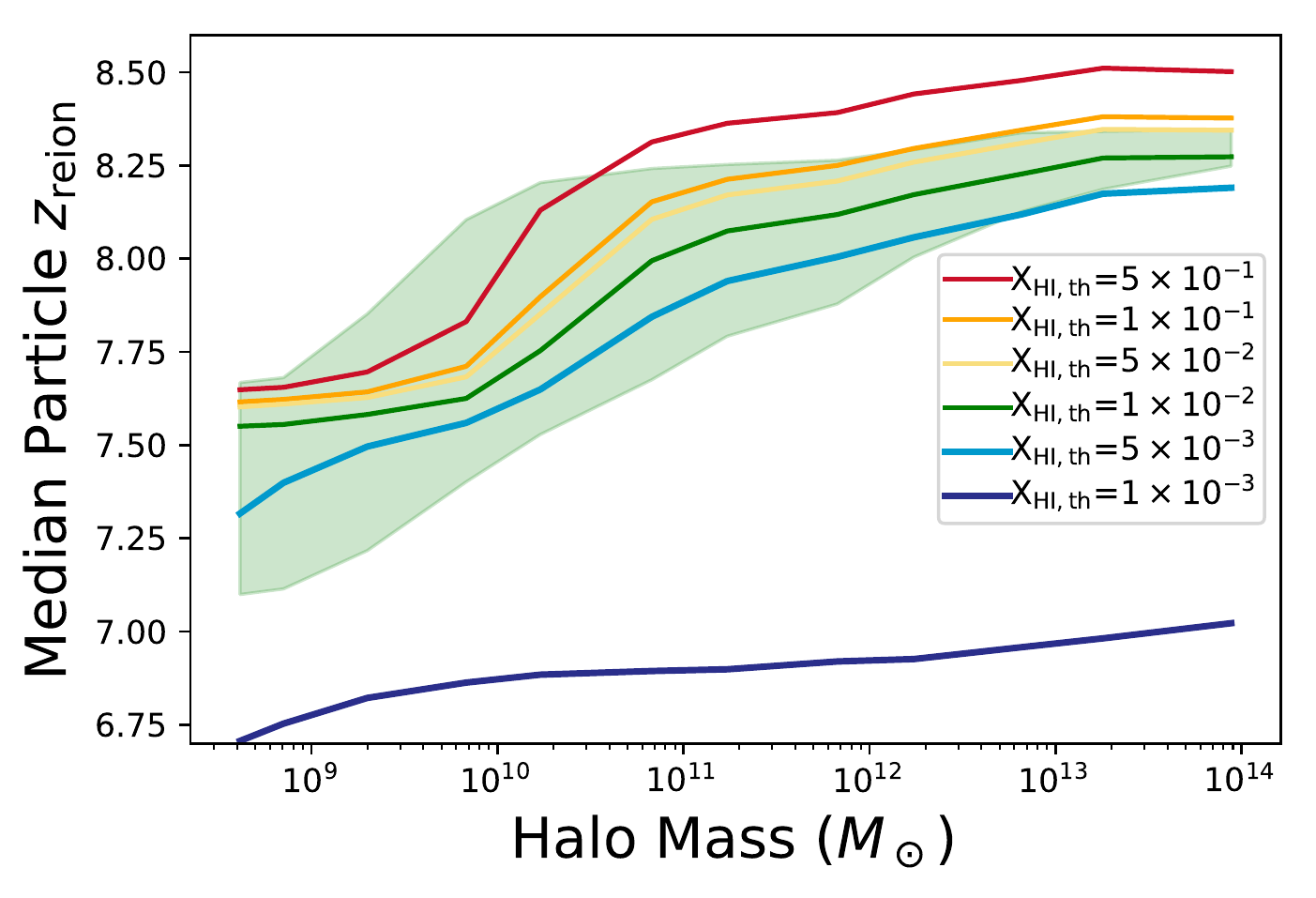} 
\end{minipage}
\caption{Halo reionization histories of our sample. Line colors correspond to different $\xHIth$. The more massive halos on average reionize earlier for all thresholds.
\label{fig:particle_median_reion_times_and_median_history}}
\end{figure}

Reionization plays a key role in galaxy formation at high redshifts, and may affect some of the properties of present-day galaxies. The simplest quantitative description of the role of reionization in the history of a given galaxy is the time the galaxy was reionized, i.e.\ its reionization redshift. We show the reionization redshift as a function of present day halo mass in Figure \ref{fig:particle_median_reion_times_and_median_history}.  Here, we use the first definition of halo reionization redshift described in Section \ref{sec:methods:reionization_redshift}, which is the median reionization redshift of the individual particle reionization redshifts from each halo.  

The solid lines in Figure~\ref{fig:particle_median_reion_times_and_median_history} correspond to the median halo reionization redshift in each mass bin.  Each color corresponds to a different neutral fraction threshold, $\xHIth$, which determines when a tracer particle is ``reionized." The shaded green region corresponds to the 25th and 75th percentile of halo reionization redshifts per mass bin for our chosen fiducial threshold, $\xHIth=0.01$.  The width of the 25-75th percentiles for other definitions is comparable.

For our fiducial threshold, halos with masses in our lowest mass range have median halo reionization redshifts of $z_{\rm reion,halo}\approx7.4$.  The median reionization redshift increases for halos in the range $10^{11}\Msun<M<10^{12}\Msun$ whose reionization redshifts scatter between $7.5\lsim z_{reion,halo}\lsim 8.3$.  The reionization redshift continues to increase with halo mass, and the shaded region has comparable width in all of our mass bins.  \emph{On average, more massive halos reionize earlier than less massive halos, regardless of the selected neutral fraction threshold for reionization.}

As we increase (decrease) the neutral fraction threshold, the halo reionization redshifts increase (decrease).  For the highest threshold we show, $\xHIth={0.5}$, the lowest mass bins have median halo reionization redshift of $z_{\rm reion,halo}\approx{7.7}$, and for the lowest threshold, $\xHI=0.001$, the corresponding reionization redshift drops to $z_{\rm reion,halo}\approx6.8$. On the other end of the halo mass bins, the highest and lowest thresholds, respectively, give $z_{\rm reion,halo}\approx{8.5}$ and $z_{\rm reion,halo}\approx7$.  Here we see that different definitions of a neutral fraction threshold for a reionized halo gives different reionization redshifts.  \emph{This indicates that the reionization process for halos is neither rapid nor instantaneous.}

Note that our highest neutral fraction threshold is the same as the single definition used in \citet{aubert_etal18}.  This results in a consistent reionization redshift in both high and low halo mass bins.  At low mass bins, our simulation results in $z_{\rm reion,halo}\approx7.7$, where their work resulted in $z_{\rm reion,halo}\approx7.8$.  At high mass bins, our simulation results in $z_{\rm reion,halo}\approx$8.5, and their work resulted in $z_{\rm reion,halo}\approx$9.

\begin{figure}
    \centering
    \includegraphics[width=\columnwidth]{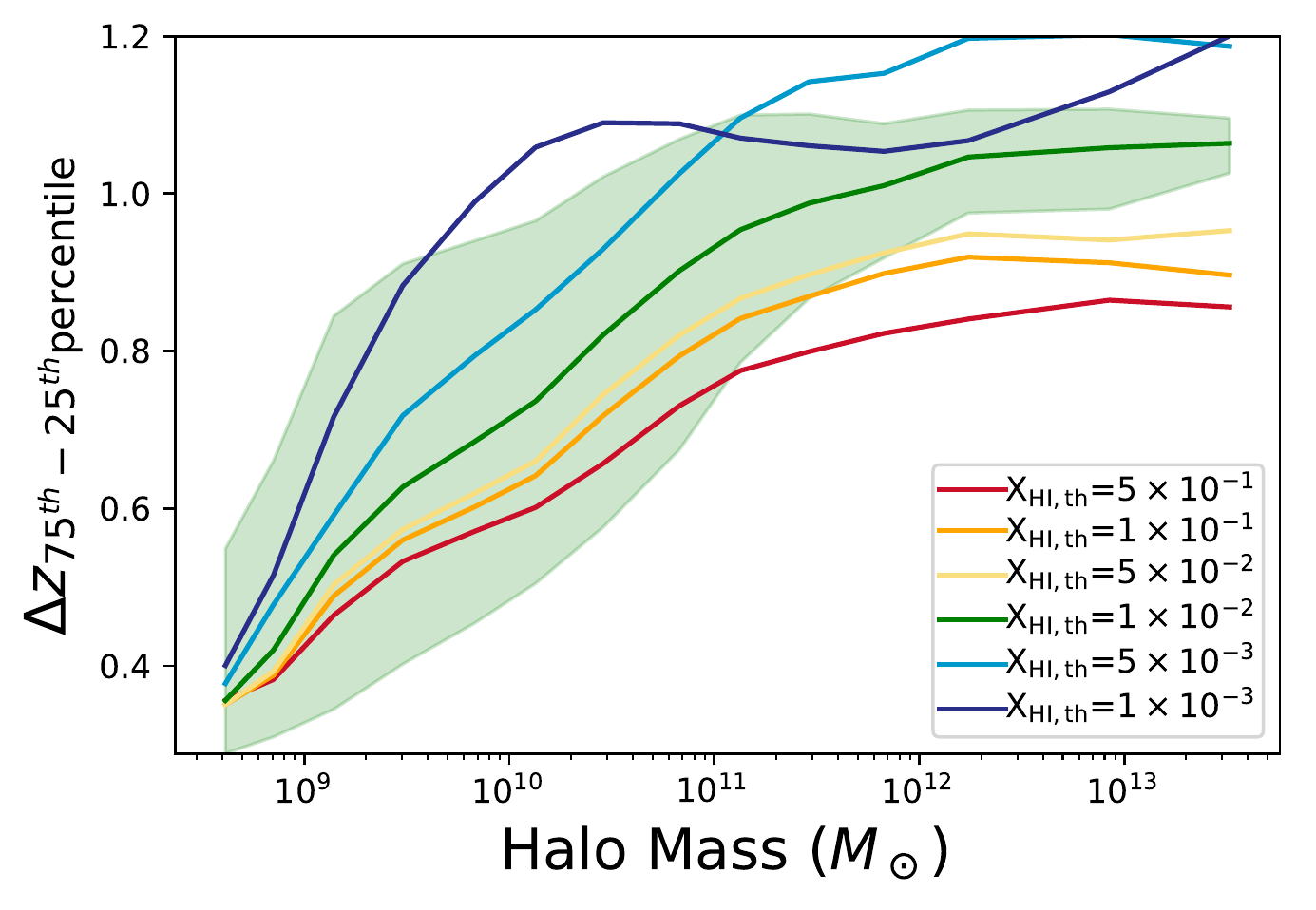}
\caption{Spread of the particle reionization times in each halo. Line colors correspond to the same $\xHIth$ as in Figure~\ref{fig:particle_median_reion_times_and_median_history}.  The smallest threshold of $\xHIth=1\times10^{-3}$ has the largest spread in particle reionization time, and the spread increases with halo mass.}
\label{fig:particle_spread_reion_times}
\vspace{5pt}
\end{figure}

Since our halo reionization histories incorporate the associated reionization history of all dark matter particles within the virial radius, we can capture the full variation of particle reionization redshifts within each halo.  We illustrate this variation in Figure~\ref{fig:particle_spread_reion_times}.  Here, we show the difference between the 75th and 25th percentiles of particle reionization redshifts in each halo, $\Delta z_{75\%-25\%}$.  We plot this spread in halo mass bins where the solid line corresponds to the median spread per mass bin, and the shaded region the 25th-75th percentile of the spread per mass bin. 

For all neutral fraction thresholds, the spread increases as a function of halo mass. More massive halos collapse from a larger volume at high redshifts, and therefore have building blocks that experience more diverse reionization histories.  

For the fiducial neutral fraction threshold, $\xHIth=0.01$ in green, the spread increases from $\Delta z_{75\%-25\%}\approx0.3$ to $\Delta z_{75\%-25\%}\approx1.0$ across the first four decades in mass. This rapid increase in spread holds for other neutral fraction thresholds that define the reionization redshift. 

On average, lower neutral fraction thresholds exhibit larger spreads. At the lowest mass bins, the chosen neutral fraction threshold does not impact the spread much. {\it Our analysis shows that the progenitors of any given present-day halo has varying reionization histories and the halo reionization history cannot be characterized only by its center or mainline progenitor.}

\section{Summary and Discussion}

We have used the cosmological simulations from the CROC project that self-consistently follow reionization processes by the UV radiation from stars, to study the reionization history of present day galaxies whose mass range lies between $10^8\Msun<M_{halo}<10^{14}\Msun$.  We explore the reionization histories of present day halos by tracing the associated reionization histories with all dark matter particles associated with the $z=0$ halo.  We find that present day halos with higher masses have earlier reionization redshifts, consistent with previous studies.  Our work shows that progenitors of our $z=0$ halos contain ``always-neutral" particles in overdense regions that are less impacted by reionizing photons, remaining substantially neutral even during redshifts when the universe is considered to be fully reionized.  Finally, our work extends previous studies by examining the spread of reionization redshifts associated with the building blocks of each halo, finding that the progenitors of higher mass halos have more diverse reionization histories leading to a larger spread of particle reionization redshifts.

Our main results are:
\begin{itemize}
\item Halo reionization process is neither rapid nor instantaneous. Varying neutral fraction thresholds for a ``reionized" halo yields different reionization redshifts.
\item Particles in the same halo have different reionization histories with a non-negligible spread in halo reionization redshifts. Therefore, we cannot simply characterize halo reionization history by tracing the halo center or its main progenitor.
\item There is a population of dark matter particles that trace dense gas regions associated with progenitor halos that stay relatively neutral through the duration of reionization.  
\end{itemize}

In future work, we plan to examine the reionization histories of different components of present day halos with a more careful analysis of progenitors connected via a merger tree.

\acknowledgments
The authors thank Huanqing Chen and Philip Mansfield for their helpful insights and comments during this project. HZ acknowledges support from the Heising-Simons Foundation, the Kavli Institute for Cosmological Physics and the Blue Waters Student Internship Program. CA acknowledges support from both the Enrico Fermi Institute and the Kavli Institute for Cosmological Physics. NG's contribution to this work used the resources of the Fermi National Accelerator Laboratory (Fermilab), a U.S. Department of Energy, Office of Science, HEP User Facility. Fermilab is managed by Fermi Research Alliance, LLC (FRA), acting under Contract No. DE-AC02-07CH11359. This work was partly supported by a NASA ATP grant NNX17AK65G, and used resources of the Argonne Leadership Computing Facility, which is a DOE Office of Science User Facility supported under Contract DE-AC02-06CH11357. An award of computer time was provided by the Innovative and Novel Computational Impact on Theory and Experiment (INCITE) program. This research is also part of the Blue Waters sustained-petascale computing project, which is supported by the National Science Foundation (awards OCI-0725070 and ACI-1238993) and the state of Illinois. Blue Waters is a joint effort of the University of Illinois at Urbana-Champaign and its National Center for Supercomputing Applications. 

\bibliographystyle{apj}
\bibliography{main}

\end{CJK*}
\end{document}